\documentstyle[12pt]{article}
\begin{document}
\pagestyle{empty}
\def\noi{\noindent}
\def\nn{\nonumber}
\def\bea{\begin{eqnarray}}  \def\eea{\end{eqnarray}}
\def\beq{\begin{equation}}   \def\eeq{\end{equation}}
\def\lsim{\raise0.3ex\hbox{$<$\kern-0.75em\raise-1.1ex\hbox{$\sim$}}}
\def\gsim{\raise0.3ex\hbox{$>$\kern-0.75em\raise-1.1ex\hbox{$\sim$}}}
\centerline{\Large \bf New CP observables in}
\vskip 3 truemm  
\centerline{\Large \bf ${\bf B}^{\bf 0}{\bf (t)} \to$ hyperon + antihyperon}
\vskip 3 truemm  
\centerline{\Large \bf from parity violation in the sequential decay} 

\vskip 8 truemm
\centerline{\bf J. Charles, A. Le Yaouanc, L. Oliver, O. P\`ene and J.-C. Raynal}
\centerline{Laboratoire de Physique Th\'eorique et Hautes 
Energies\footnote{Laboratoire associ\'e
au Centre National de la Recherche Scientifique - URA D0063}} 
\centerline{Universit\'e de Paris XI, B\^atiment 210,
F-91405 Orsay Cedex, France}

\vskip 3 truemm 
\begin{abstract}
We consider the decay $B^0(t) \to$ hyperon + antihyperon, followed by hyperon weak
decay. We show that parity violation in the latter allows to reach new CP 
observables~: not only
Im$\lambda_f$ but also Re$\lambda_f$ can be measured. In the decay $B^0_d(t) \to 
\Lambda
\bar{\Lambda}$ $(BR \sim 10^{-6})$, $\Lambda \to p\pi^-$ these observables reduce 
to $\sin 2\alpha$
and $\cos 2\alpha$ in the small Penguin limit, the latter solving the discrete 
ambiguity $\alpha
\to {\pi \over 2} - \alpha$. For $\beta$ one could consider the Cabibbo suppressed 
mode $B^0_d(t)
\to \Lambda^+_c \overline{\Lambda^+_c}$ $(BR \sim 10^{-4})$, $\Lambda^+_c \to 
\Lambda \pi^+$,
$p\bar{K}^0$, ... (with $BR \sim 10^{-2})$. The pure Penguin modes $B^0_s(t) \to
\Sigma^-\overline{\Sigma^-}$, $\Xi^-\overline{\Xi^-}$, 
$\Omega^-\overline{\Omega^-}$ $(BR \sim
10^{-7})$ can be useful in the search of CP violation beyond the Standard Model. 
Because of the
small total rates, the study of these modes could only be done in future high 
statistics
experiments. Also, in the most interesting case $\Lambda\bar{\Lambda}$ the time 
dependence of the
asymmetry can be difficult to reconstruct. On the other hand, we show that $B_d$ 
mesons, being a
coherent source of $\Lambda\bar{\Lambda}$, is useful to look for CP violation in 
$\Lambda$
decay. We also discuss $B^0_d(t) \to J/\Psi K^{*0} \to \ell^+ \ell^- K_S\pi^0$ 
where the secondary
decays conserve parity, and angular correlations allow to determine terms of the 
form $\cos \delta
\cos 2\beta$, $\delta$ being a strong phase. This phase has been measured by CLEO, 
but we point out
that a discrete ambiguity prevents to determine sign($\cos 2\beta$). However, if 
one assumes small
strong phases, like in factorization and as supported by CLEO data, one could have 
information
on sign($\cos 2\beta$). Similar remarks can be done for $\cos 2 \alpha$ in the 
decay $B_d^0(t) \to
\rho \rho \to 4 \pi$.   \end{abstract} 

\noindent LPTHE Orsay 98-21 \par
\noindent March 1998 \par
\newpage
\pagestyle{plain}
\baselineskip=24 pt

\section{Introduction}
\hspace*{\parindent} 
The measurement of the CP angles of the Unitarity Triangle (UT) through
time-dependent CP asymmetries is a major purpose of particle physics in the next 
years.
 
	Concerning the angle $\alpha$, the present determination of the sides of 
the UT
gives a range $40^{\circ} < \alpha < 140^{\circ}$ \cite{1r}. The CLEO upper bound 
on the favorite
decay $B_d \to \pi^+\pi^-$ \cite{2r}, although not inconsistent with naive 
expectations, points to
a determination of $\alpha$ that will not be easy. The presence of Penguins 
complicates the picture,
mostly if the expected color suppression of the $\pi^0\pi^0$ mode does not allow 
to
perform an isospin analysis \cite{3r}, \cite{4r}.  Therefore, it may happen that 
one will only
get an effective angle $\alpha_{eff}$, which is related to $\alpha$ via the 
Penguin
contribution \cite{5bisr}. Also, even if $\sin 2\alpha_{eff}$ is some day measured 
in
$B^0_d(t) \to \pi^+\pi^-$, the discrete ambiguity $\alpha_{eff} \to {\pi \over 2}  
- \alpha_{eff}$
(and also $\pi + \alpha_{eff}$) would be left \cite{5r}. The ambiguity 
$\alpha_{eff} \to {\pi \over
2}  - \alpha_{eff}$ could be solved by the measurement of sign($\cos 
2\alpha_{eff}$). The
measurement of $\sin 2\alpha$, $\cos 2\alpha$, and of the Penguin amplitudes could 
be made in
principle by the study of the time-dependent Dalitz plot $B^0_d(t) \to 
\pi^+\pi^-\pi^0$ via $\rho$
decay \cite{6r}. However, the expected branching ratios for the different $\rho 
\pi$ decay modes
make this study difficult, at least in the first generation of CP violation 
experiments in B mesons.
It is therefore suitable to study all possible decay modes that can help to give 
hints on the
different aspects of the measurement of $\alpha$. In this paper we propose to 
consider the
sequential decay $B^0_d(t) \to \Lambda \bar{\Lambda}$, $\Lambda \to p\pi^-$ that 
allows in principle
to measure both $\sin 2\alpha$ and also $\cos 2\alpha$ (up to Penguins) thanks to 
parity violation
in $\Lambda$ decay. The new CP information is on the sign of $\cos 2 
\alpha_{eff}$, where
$\alpha_{eff}$ is related to $\alpha$ via the Penguin contributions \cite{5bisr}. 
The relevant
features of this decay, with an amplitude $A(B^0_d \to \Lambda \bar{\Lambda}) \sim 
|V_{ub}|$ and
hence an expected branching ratio of the order of $10^{-5}$-$10^{-6}$, the 
excellent detection
efficiency of $\Lambda \to p\pi^-$ with a large $BR(\Lambda \to p\pi^-) = 64 \ 
\%$, and a sizeable
parity violation parameter $\alpha (p\pi^-) = 0.64$, necessary to get information 
on $\cos
2\alpha_{eff}$ as we will show below, make this mode very interesting. Let us 
emphasize however again
that this mode gets contributions not only from current-current operators but also 
from local Penguin
operators  (Fig.~1), and long distance Penguin contributions as well 
\cite{18bisr}. 

The decay mode $B^0_d(t) \to J/\Psi K_S$ will hopefully allow to measure $\sin 
2\beta$ with a high
precision. Considering the decay modes studied in this paper, the same type of 
arguments can be
applied to the angle $\beta$ through the Cabibbo-supressed decay $B^0_d(t) \to 
\Lambda^+_c
\overline{\Lambda^+_c}$ (Fig.~2) since $\Lambda^+_c$ decays weakly, like in 
$\Lambda^+_c \to
\Lambda \pi^+$, $p\bar{K}^0$, ... These sequential decays could in principle allow 
to measure
$\sin 2\beta$ and also $\cos 2\beta$ up to possible Penguin pollution.

For $\gamma$ one could naively consider $B^0_s(t) \to \Lambda \bar{\Lambda}$,
$\Xi^0\overline{\Xi^0}$,  but branching ratios are expected to be very small in 
this case and
Penguins are large (Fig.~3), as we will see below.
 
We also point out that the pure Penguin modes $B^0_s(t) \to 
\Sigma^-\overline{\Sigma^-}$,
$\Xi^-\overline{\Xi^-}$, $\Omega^-\overline{\Omega^-}$ (Fig.~4), for which the 
Standard Model
predicts very small asymmetries, could be useful in the search of CP violation 
beyond the Standard
Model. Moreover, these pure Penguin modes can give a hint on the strength of the 
possible Penguin
pollution in the decays relevant for $\alpha$, $\beta$ or $\gamma$. 

Let us first estimate the order of magnitude of the branching ratios of the modes 
that we 
discuss in this paper. For $B_d^0 \to \Lambda_c^+ \overline{\Lambda_c^+}$ we can 
make a very
crude estimate from the measured inclusive ratio $BR(B^{\pm}/B^0 \to 
\bar{\Sigma}_c +
\hbox{anything}) \cong 5 \times 10^{-3}$ \cite{7r}~: 

\beq
BR (B_d^0 \to \Lambda_c^+ \overline{\Lambda_c^+}) \sim {BR (\bar{\Sigma}_c + 
\hbox{anything})
\over BR(D + \hbox{anything})} \times BR(\bar{D}D_s) \sim 2 \times 10^{-4}
\label{1ebis}
 \eeq

\noindent and for $B_d^0 \to \Lambda\bar{\Lambda}$ we rescale from the ratio 
$\left | {V_{ub} \over
V_{cb}} \right | \cong 0.08$~:

\beq
 BR (B_d^0 \to \Lambda \bar{\Lambda}) \sim 10^{-6} \quad . 
\label{2ebis}
\eeq

For $B_s \to \Lambda \bar{\Lambda}$, $\Xi^0\overline{\Xi^0}$ there is a further 
Cabibbo
suppression, yielding very small $BR \sim 10^{-7} - 10^{-8}$. The pure Penguin 
modes $B_s^0 \to
\Sigma^- \overline{\Sigma^-}$, $\Xi^-\overline{\Xi^-}$, $\Omega^- 
\overline{\Omega^-}$ can be
rescaled from $BR (B_d^0 \to K\pi ) \sim 10^{-5}$ yielding $BR \sim 10^{-7}$, 
while the branching
ratios of modes of $B_d^0$ to the same final states, that are also of Penguin 
type, are predicted to
be very tiny, $BR \sim 10^{-8} - 10^{-9}$ \cite{par6}.

Let us first discuss the angle $\beta$ and the well known angular correlations 
$B^0_d(t)
\to J/\Psi K^{*0} \to \ell^+\ell^-K_S\pi^0$ \cite{8r} \cite{9r}, where the 
secondary decays conserve
parity and Penguin pollution is expected to be very small. This will make a 
natural introduction to
the new angular correlations that appear if the secondary decay violates parity. 
	
In the Standard Model, by the measurement of the sides of the unitarity triangle, 
$\beta$ is
constrained to be in the range $10^{\circ} \leq \beta \leq 35^{\circ}$ \cite{1r}. 
Measurement of
$\sin 2\beta$ in this range in e.g. $B^0_d(t) \to J/\Psi K_S$ would leave the 
ambiguities $\beta \to
{\pi \over 2}  - \beta$, $\pi + \beta$, that would correspond, owing to the range 
predicted by the
Standard Model, to possible physics beyond the Standard Model. Recently, some 
theoretical effort has
been devoted to this question of the discrete ambiguities \cite{5r}. Concerning 
the ambiguity
$\beta \to {\pi \over 2}  - \beta$, that needs the measurement of sign($\cos 
2\beta$), we have put
forward some ideas, like Dalitz plot analyses like $B^0_d(t)  \to D^+D^-K_S$ via
$D_s^{**}$ and other possible channels \cite{10r}. Another nice proposal involves 
interference
between $B^0_d$-$\bar{B}^0_d$ mixing and $K^0$-$\bar{K}^0$ mixing in the cascade 
decay
$B^0_d(\bar{B}^0_d)  \to J/\Psi K \to J/\Psi \pi \ell \nu$ \cite{11r}. 
	
As we will see in detail below, in the time-dependent decay $B^0_d(t) \to
J/\Psi K^{*0} \to \ell^+\ell^-K_S\pi^0$ the observables obtained from angular 
analysis are 
\cite{9r} $|G_0(t)|^2$, $|G_+(t)|^2$, $|G_-(t)|^2$, Re$[G_0(t)G^*_+(t)]$, 
Im$[G_+(t)G^*_-(t)]$ and
Im$[G_0(t)$ $G^*_-(t)]$, where $G_0(t)$, $G_+(t)$, $G_-(t)$ are transversity 
amplitudes to final
states of definite CP (respectively $+$, $+$, $-$) \cite{12r}. The observables of 
the form
Im$[G_{CP=+}(t)$ $G^*_{CP=-}(t)]$ contain the term Re$[G_{CP=+}(0)G^*_{CP=-}(0) 
]\cos 2
\beta$\break \noindent $\sin \Delta Mt$, and the determination of sign($\cos 
2\beta$), necessary to
lift the ambiguity $\beta \to {\pi \over 2}  - \beta$, is polluted by a 
coefficient $\cos \delta$,
$\delta$ being the strong phase $\delta = {\rm arg}[G_{CP=+}(0)G^*_{CP=-}(0) ]$ 
\cite{13r}. This
phase has been recently measured by CLEO \cite{14r}, but, as we will show below 
there is a discrete
ambiguity on $\delta$ that does not allow to measure sign($\cos 2\beta$) in a 
time-dependent
analysis, unless some reasonable model-dependent assumption is made. We will show 
that if one
assumes small strong phase shifts in the decay $B^0_d  \to J/\Psi K^{*0}$ one 
could determine
sign($\cos 2\beta$) modulo this hypothesis. This is en\-cou\-ra\-ging because CLEO 
finds amplitudes
that are close to be relatively real, as predicted by factorization.  

The difficulty of the pollution by strong phases in the determination of $\cos 
2\beta$ is lifted in
principle if one considers the Cabibbo suppressed decay $B^0_d(t) \to \Lambda^+_c
\overline{\Lambda^+_c}$. The baryon $\Lambda^+_c$ decaying weakly, e.g. 
$\Lambda^+_c \to \Lambda
\pi^+$, $p\bar{K}^0$, ... we will show below that other observables are accessible 
because of parity
vio\-la\-tion. Calling $G_+(t)$, $G_-(t)$ the amplitudes to $\Lambda^+_c 
\overline{\Lambda^+_c}$ in
a definite CP state, the observables are now $|G_+(t)|^2$, $|G_-(t)|^2$, 
Re$[G_+(t)G^*_-(t)]$,
Im$[G_+(t)G^*_-(t)]$, allowing to measure $\sin 2\beta$ and also $\cos 2\beta$ (in 
the small
Penguin limit) without strong phase pollution. The same argument applies to the 
more interesting case
of $\alpha$, namely the sequential decay $B^0_d(t) \to \Lambda \bar{\Lambda}$, 
$\Lambda \to p\pi^-$,
the main result of this paper. Moreover, we emphasize that this decay could be 
useful to look for
CP violation in $\Lambda$ decay.\\

\section{Remarks on angular analysis in B$_{\bf d}^{\bf 0} \to {\bf J/}\Psi\ {\bf 
K^{*0}}$}
\hspace*{\parindent} 
To introduce the subject, let us first discuss the well-known case $B^0_d(t) \to 
J/\Psi$ $K^{*0} \to
\ell^+\ell^- K_S\pi^0$. As explained in detail in ref. \cite{9r}, the angular 
dependent rate takes
the form

\beq
|M|^2 = \left ( {3 \over 4\pi}\right )^2 \sum_{\alpha = \pm 1}  \left | 
\sum_{\lambda =0,\pm 1} 
A_{\lambda} \  D^1_{\lambda ,\alpha}\left ( R_{\Psi} \right )^* D^1_{\lambda 
,0}\left ( R_{K^*}
\right )^* \right |^2 	  \label{1e} \eeq

\noindent where $A_{\lambda}$ are helicity amplitudes, in general 
\underbar{time-dependent}~: the
angular dependence and the time dependence factorize. For the Jackson convention, 
$R = (\theta ,
\varphi ,0)$ where $\theta$, $\varphi$ are the decay angles in the vector meson 
rest frame. $|M|^2$
can be written in the form   

\bea
&&|M|^2 = \left ( {3 \over 4\pi} \right )^2 \sum_{\alpha = \pm 1} \sum_{\lambda = 
0, \pm 1}
\sum_{\lambda '=0, \pm 1}  A_{\lambda} A^*_{\lambda '}  (-1)^{2\lambda - \alpha} 
\sum_{J_L=0,1,2} 
\sum_{J_R=0,1,2} \nn \\  
&&< 1 \alpha , 1 - \hskip -1 truemm \alpha |J_L 0 > < 1 \lambda ' , 1 - \hskip -1 
truemm \lambda
|J_L \lambda '\hskip -1 truemm - \hskip -1 truemm \lambda > D^{J_L}  _{\lambda '- 
\lambda
,0}(R_{\Psi}) \nn \\  
&&< 1 0 , 1 0|J_R 0 > < 1 \lambda ' , 1 - \hskip -1 truemm \lambda |J_R \lambda 
'\hskip -1 truemm -
\hskip -1 truemm \lambda > D^{J_R}_{\lambda '- \lambda ,0}(R_{K^*}) \nn \\  
&&= \left ({3 \over
4\pi}\right )^2 \sum_{\alpha = \pm 1} \sum_{\lambda =0,\pm 1} \sum_{\lambda '=0, 
\pm 1}  A_{\lambda}
A^*_{\lambda '}  (-1)^{2\lambda - \alpha} \sum_{J_L=0,1,2}  \sum_{J_R=0,1,2} \nn 
\\    
&&< 1 \alpha
, 1 - \hskip -1 truemm \alpha |J_L 0 > < 1 \lambda ' , 1 - \hskip -1 truemm 
\lambda |J_L \lambda '\hskip -1 truemm
- \hskip -1 truemm \lambda > \nn \\   
&&< 1 0 , 1 0|J_R 0 > < 1 \lambda ' , 1 - \hskip -1 truemm \lambda |J_R \lambda
'\hskip -1 truemm - \hskip -1 truemm \lambda > \nn\\  &&\sqrt{{4\pi \over 2J_L+1}} 
 \sqrt{{4\pi
\over 2J_R+1}}   Y^* _{J_L \lambda '-\lambda}(\Omega_{\Psi})  Y^*_{J_R \lambda
'-\lambda}(\Omega_{K^*}) \quad . 				 \label{2e} \eea

\noindent Taking moments

\beq
T_{J_LJ_RM} = \int   |M|^2 Y_{J_L M}(\Omega_{\Psi}) \ Y_{J_R M}(\Omega_{K^*}) 
d\Omega_{\Psi}
d\Omega_{K^*}		 \label{3e} \eeq

\noi one finds 
\newpage
\bea
&&T_{J_LJ_RM} =  \left ( {3 \over 4\pi} \right )^2 \sqrt{{4\pi \over 2J_L+1}}  
\sqrt{{4\pi
\over 2J_R+1}}  \sum_{\alpha = \pm 1} \sum_{\lambda =0, \pm 1} \sum_{\lambda '=0, 
\pm 1} 
A_{\lambda} A^*_{\lambda '}  (-1)^{2\lambda - \alpha} \nn \\ 
&&< 1 \alpha , 1- \hskip -1 truemm \alpha |J_L 0 > < 1
\lambda ' , 1-\hskip -1 truemm\lambda |J_L M > \nn \\  
&&< 1 0 , 1 0|J_R 0 > < 1 \lambda ' , 1-\hskip -1 truemm \lambda |J_R M
>				 \label{4e} \eea

\noindent with the relation 

\beq
			T^* _{J_LJ_RM}  = T_{J_LJ_R-M} \quad .						
\label{5e}
\eeq

\noindent In terms of transversity amplitudes 

\beq
		G_0 = A_0	\qquad G_+ = {A_{+1}+A_{-1} \over \sqrt{2}} \qquad 	
G_- = {A_{+1} - A_{-1} \over
\sqrt{2}}  		 \label{6e}
\eeq

\noindent one finds that the non-vanishing moments are the following \cite{9r}~:

\bea
&&T_{000} = {2 \over 4\pi} \left ( |G_0|^2 + |G_+|^2 + |G_-|^2 \right ) \nn \\ 
&&T_{020} = {2 \over 4\sqrt{5} \pi} \left ( 2|G_0|^2 - |G_+|^2 - |G_-|^2 \right ) 
\nn \\
&&T_{200} = {1 \over 4 \sqrt{5} \pi} \left (- 2|G_0|^2 + |G_+|^2 + |G_-|^2 \right 
) \nn \\
&&T_{220} = - {1 \over 5} \left ( 4|G_0|^2 + |G_+|^2 + |G_-|^2 \right ) \nn \\ 
&&T_{22-1} = - {3 \sqrt{2} \over 5} \left ( {\rm Re} G_+G^*_0 + i{\rm Im}G_-G^*_0 
\right ) \nn \\ 
&&T_{22-2} = - {3 \over 5} \left ( |G_+|^2 - |G_-|^2 + 2i{\rm Im}G_-G^*_+ \right ) 
\quad . 						
\label{7e}
\eea

\noindent We have used the notation $G_+$ and $G_-$ instead of $G_{1+}$ and 
$G_{1-}$ of
ref.~\cite{9r} for the transverse CP even and transverse CP odd amplitudes to make 
explicit the
differences with the $\Lambda^+_c \overline{\Lambda^+_c}$ case, where we will use 
the same notation
$G_+$, $G_-$ for the CP even and CP odd amplitudes. We see that the observables 
are
 
\beq
|G_0|^2     \qquad      |G_+|^2    \qquad        |G_-|^2	   \qquad   {\rm 
Re} G_+G^*_0  	 \qquad  
{\rm Im} G_-G^*_0  	  \qquad    {\rm Im}G_-G^*_+ 	    \label{8e}
\eeq

\noindent that are in general time dependent. One can measure the \underbar{real 
parts} of the
interferences between amplitudes of \underbar{same CP} and the \underbar{imaginary 
parts} of the
interferences of amplitudes of \underbar{opposite CP}.  

The time dependence of these amplitudes follows from the
expression for the time evolution ($\Delta \Gamma \cong 0$ and $\left |{q \over p} 
\right | = 1$ is
assumed, as given by the Standard Model)~: 

\bea
&&|B^0_d(t)  > = e^{-iMt} \  e^{-\Gamma t/2} \left ( \cos {\Delta Mt \over 2} 
|B^0_d > + \ i {q \over
p} \sin {\Delta Mt \over 2} |\bar{B}^0_d > \right )\nn \\
&&|\bar{B}^0_d(t)  > = e^{-iMt} e^{-\Gamma t/2} \left ( \cos {\Delta Mt \over 2} 
|\bar{B}^0_d > + \
i {p \over q} \sin {\Delta Mt \over 2} |B^0_d > 	\right )	 
\label{9e}
\eea

\noindent that gives, for $B^0_d(t)$ decay amplitudes

\bea
&&G_0(t) = G_0(0) \ e^{-iMt} e^{-\Gamma t/2} \left ( \cos {\Delta Mt \over 2}  + 
i\eta_f  \
e^{-2i\beta} \sin {\Delta Mt \over 2} \right ) \nn \\ 
&&G_+(t) = G_+(0) \ e^{-iMt} e^{-\Gamma t/2} \left ( \cos {\Delta Mt \over 2}  + 
i\eta_f \ 
e^{-2i\beta} \sin {\Delta Mt \over 2} \right ) \nn \\ 
&&G_-(t) = G_-(0) \ e^{-iMt} e^{-\Gamma t/2} \left ( \cos {\Delta Mt \over 2}  - 
i\eta_f
\ e^{-2i\beta} \sin {\Delta Mt \over 2} \right ) 			 
\label{10e}
\eea

\noindent and analogously for $\bar{B}^0_d(t)$ decays.  In these expressions, the 
sign $\eta_f$
depends on the CP eigenstate in which $K^{*0}$ decays, for example $\eta (K_S 
\pi^0) = +$. Notice
that in these expressions we have neglected possible Penguin contributions, that 
are expected to
be very small in these $J/\Psi K^*$ modes. 

 Then, the time dependent observables write,  

\bea
&&|G_0(t)|^2 = |G_0(0)|^2\ e^{-\Gamma t} \left (1 + \eta_f \sin 2\beta \sin \Delta 
Mt \right )  \nn
\\ &&|G_+(t)|^2 = |G_+(0)|^2\ e^{-\Gamma t} \left (1 + \eta_f \sin 2 \beta \sin 
\Delta Mt \right) \nn
\\ &&|G_-(t)|^2 = |G_-(0)|^2\ e^{-\Gamma t} \left (1 - \eta_f \sin 2\beta \sin 
\Delta Mt \right ) \nn
\\ &&{\rm Re} \left [ G_+(t)G^*_0(t) \right ]  = {\rm Re} \left [ G_+(0)G^*_0(0) 
\right ] 
e^{-\Gamma t} \left (1 + \eta_f  \sin 2\beta \sin \Delta Mt \right )	\nn \\	 
\label{11e}
&&{\rm Im} \left [ G_-(t)G^*_0(t) \right ] = e^{-\Gamma t} \Big \{ - {\rm Re} 
\left [
G_-(0)G^*_0(0) \right ] \eta_f \cos 2\beta \sin \Delta Mt + \nn \\
&& \qquad  {\rm Im} \left [ G_-(0)G^*_0(0) \right ]\cos \Delta Mt \Big \} \nn \\
&&{\rm Im} \left [ G_-(t)G^*_+(t) \right ]  = e^{-\Gamma t} \Big \{ - {\rm Re} 
\left [
G_-(0)G^*_+(0) \right ] \eta_f \cos 2\beta \sin \Delta Mt + \nn \\
&& \qquad {\rm Im} \left [ G_-(0)G^*_+(0) \right ] \cos \Delta Mt \Big \}   \eea

\noi CLEO has measured these observables in a time-integrated experiment 
\cite{14r}, that amounts
to determine these quantities at $t = 0$.

>From these expressions we see that in a time-dependent angular a\-na\-ly\-sis 
experiment where we
could hopefully separate the time dependence $e^{-\Gamma t} \sin \Delta Mt$, one 
could measure the
products 

\beq
{\rm 	Re} \left [ G_-(0)G^*_0(0) \right ] \cos 2\beta \qquad 	{\rm	Re}\left [ 
G_-(0)G^*_+(0)
\right ] \cos 2\beta \quad .		 \label{12e}
\eeq

\noi Therefore, one can only measure products of the form $\cos \delta \cos 
2\beta$ where $\delta$ is
the strong phase Arg$[G_-(0)G^*_0(0)]$ or Arg$[G_-(0)G^*_+(0)]$. Then, this 
measurement could not
solve the ambiguity $\beta \to {\pi \over 2}  - \beta$ by itself because of the 
strong phase
$\delta$ \cite{13r}. However, as we will see below, there is some experimental 
knowledge on these
phases that could allow to have information on $\cos 2\beta$ if some additional 
theoretical input is
assumed \cite{14bisr}.

Let us go back to the decay $J/\Psi (K^{*0})_{K_S\pi^0}$ and to the CLEO data 
\cite{14r}. CLEO has
reported the phases, within the convention $\varphi (G_0) = 0$~:

\beq
\varphi (G_+) + \pi = 3.00 \pm 0.37 \pm 0.04		\qquad \varphi (G_-) = 
-0.11 \pm 0.46 \pm 0.03	
\label{13e}
\eeq

\noindent (the $\pi$ comes from the particular CLEO convention). These results are 
consistent with
the amplitudes $G_0$, $G_+$, $G_-$ being real relatively to each other, as 
expected within the
hypothesis of factorization.  

The CLEO results seem to solve the problem of determining sign($\cos 2\beta$),
since $\varphi (G_+)$-$\varphi (G_0)$ and $\varphi (G_-)$-$\varphi (G_0)$ have 
been measured. There
is, however, a discrete ambiguity in the determinations of these phases, and 
therefore a second
solution. From the angular distribution \cite{14r}, i.e. simply the observables 
quoted
above at $t = 0$, one sees that one can measure (within the CLEO convention 
$\varphi (G_0) = 0$)~:

\beq
	\cos \varphi (G_+)		\qquad \sin [\varphi (G_-)-\varphi (G_+)]	
	\qquad \sin \varphi (G_-)	\quad .	
\label{14e}
\eeq

These quantities remain invariant under

\beq
		\varphi (G_+) \to - \varphi (G_+)	\qquad 	\varphi (G_-) \to 
\pi - \varphi (G_-)		
\label{15e}
\eeq

\noindent and there is a second solution for these phases.
 
In the time-dependent analysis, the terms proportional to $\cos 2\beta$ change 
sign under this
transformation
\bea
&&\cos \left [ \varphi (G_-) - \varphi (G_+) \right ] \cos 2\beta  \to  - \cos 
\left [
\varphi (G_-) - \varphi (G_+) \right ] \cos 2 \beta \nn \\
&&\cos \varphi (G_-) \cos 2 \beta \to - \cos \varphi (G_-) \cos 2 \beta \quad .					
\label{16e}
\eea

\noindent There is a sign ambiguity on $\cos [ \varphi(G_-)-\varphi (G_+)]$ and on 
$\cos \varphi
(G_-)$ and therefore a sign ambiguity on $\cos 2\beta$ remains \cite{14bisr}. One 
of the solutions
for sign$\{ \cos [ \varphi (G_-)-\varphi (G_+)]\}$ and on sign$[ \cos 
\varphi(G_-)]$ will correspond
to the re\-la\-ti\-ve sign between CP even and CP odd amplitudes as given by 
factorization
\cite{15r}, \cite{16r}. The other solution will correspond to the (awkward~?) 
solution in which the
relative sign has been exactly reversed by a very large FSI. Then, the measurement 
of the relative
phases plus the hypothesis of small strong phase shifts can give a hint on the 
$\beta \to {\pi \over
2}  - \beta$ ambiguity. Let us emphasize that \underbar{there are two observables 
of this kind}, and
then if sign$[\cos \delta \cos 2 \beta ]$ is observed to be the same for both, 
consistent with
$\beta$ as constrained by the Standard Model and with small strong phase $\delta$, 
we could
confidently conclude about the Standard Model solution for $\beta$. It is 
important to note that the
same remarks can be done to have information on $\cos 2 \alpha$, since observables 
of the same form
are in principle measurable in $B_d^0(t) \to \rho \rho \to 4 \pi$, although 
Penguin pollution is
expected to be sizeable in this case. \\

\section{Angular analysis in B$^{\bf 0} \to$ hyperon + antihyperon}
\hspace*{\parindent} 
Let us now turn back to the decays $B^0(t) \to$ hyperon + antihyperon, taking 
first as an example
$B_d^0(t) \to \Lambda_c^+ \overline{\Lambda_c^+}$. One can write the angular 
distribution

\beq
|M|^2 = {1 \over 4\pi^2} \sum_{\alpha = \pm 1/2}  \sum_{\beta = \pm 1/2}  \left | 
\sum_{\lambda =
\pm1/2}  A_{\lambda} B_{\alpha} \bar{B}_{\beta} D^{1/2}_{\lambda ,\alpha } \left ( 
R_{\Lambda_c}
\right )^* D^{1/2}_{\lambda ,\beta} \left ( R_{\bar{\Lambda}_c} \right )^* \right 
|^2 
\label{17e}
\eeq

\noindent where $A_{\lambda}$ are the helicity amplitudes of the decay $B^0_d  \to 
\Lambda^+_c 
\overline{\Lambda^+_c}$ in the $B$ center-of-mass and $B_{\alpha}$ 
$(\bar{B}_{\beta})$ are the
helicity amplitudes of the decay $\Lambda^+_c  \to \Lambda \pi^+$ 
$(\overline{\Lambda^+_c}  \to
\bar{\Lambda} \pi^-)$ in the $\Lambda_c(\bar{\Lambda}_c)$ rest frames. After some 
angular momentum
calculations, one finds, analogously to (\ref{2e})~:

\newpage
\bea
&&|M|^2 = {1 \over 4\pi^2} \sum_{\alpha = \pm 1/2}  \sum_{\beta = \pm1/2} 
\sum_{\lambda =\pm 1/2}
\sum_{\lambda '=\pm 1/2}  A_{\lambda} \ A^*_{\lambda '} \   B_{\alpha}  \ 
\bar{B}_{\beta} \ 
B^*_{\alpha} \   \bar{B}^*_{\beta} \nn \\
&&(-1)^{2\lambda -\alpha -\beta}  \sum_{J_L=0,1}  \sum_{J_R=0,1}   \nn \\
&&< {1 \over 2}  \alpha , {1 \over 2}  -\hskip -1 truemm \alpha |J_L 0 > <{1 \over 
2} \lambda ' , {1
\over 2}  -\hskip -1 truemm \lambda |J_L \lambda '\hskip -1 truemm - \hskip -1 
truemm \lambda > \nn
\\ &&< {1 \over 2} \beta  , {1 \over 2}  -\hskip -1 truemm \beta |J_R 0 > < {1 
\over 2}  \lambda ' ,
{1 \over 2}  -\hskip -1 truemm \lambda |J_R \lambda '\hskip -1 truemm - \hskip -1 
truemm \lambda >
\nn \\ && \sqrt{{4\pi \over 2J_L+1}}  \sqrt{{4\pi \over 2J_R+1}} \   Y^*_{J_L 
\lambda '-\lambda}\left
(\Omega _{\Lambda_c} \right ) \  Y^*_{J_R \lambda '-\lambda} \left 
(\Omega_{\bar{\Lambda}_c}
\right ) 				 \label{18e} \eea

\noindent and defining moments~:

\beq
T_{J_LJ_RM} = \int  |M|^2 \ Y_{J_L M} \left ( \Omega_{\Lambda_c} \right ) \ Y_{J_R 
M}\left (
\Omega_{\bar{\Lambda}_c} \right ) d\Omega_{\Lambda_c} \ d\Omega_{\bar{\Lambda}_c}	
		 \label{19e}
\eeq

\noindent one obtains

\bea
&&T_{J_LJ_RM} = {1 \over 4\pi^2}  \sqrt{{4\pi \over 2J_L+1}}  \sqrt{{4\pi \over 
2J_R+1}} \nn \\ 
&&\sum_{\alpha = \pm 1/2}  \sum_{\beta = \pm 1/2} \sum_{\lambda = \pm 1/2} 
\sum_{\lambda '= \pm
1/2}  A_{\lambda} \ A^*_{\lambda '} \ B_{\alpha} \ \bar{B}_{\beta} B^*_{\alpha} \ 
\bar{B}^*_{\beta} 
(-1)^{2\lambda - \alpha - \beta} \nn \\ 
&&< {1 \over 2} \alpha , {1 \over 2}  -\hskip -1 truemm \alpha |J_L 0 > < {1 \over 
2}  \lambda ' ,
{1 \over 2}  - \hskip -1 truemm \lambda |J_L M > \nn \\ 
&&< {1 \over 2}  \beta , {1 \over 2}  -\hskip -1 truemm\beta |J_R 0 > < {1 \over 
2}  \lambda ' , {1 \over 2} 
-\hskip -1 truemm \lambda |J_R M >  				 \label{20e} \eea

\noi and the same relation (\ref{5e}), $T^*_{J_LJ_RM}  = T_{J_LJ_R-M}$.

One finds the independent moments
\newpage
\bea
&&T_{000} = {1 \over 4\pi}  \left [ |A_{+1/2}|^2 +
|A_{-1/2}|^2 \right ] \left [ |B_{+1/2}|^2 + |B_{-1/2}|^2 \right ] \left [
|\bar{B}_{+1/2}|^2 + |\bar{B}_{-1/2}|^2 \right ] \nn \\
&&T_{100} = {1 \over 4\pi \sqrt{3}} \left [
|A_{+1/2}|^2 - |A_{-1/2}|^2 \right ] \left [ |B_{+1/2}|^2 - |B_{-1/2}|^2
\right ] \left [ |\bar{B}_{+1/2}|^2 + |\bar{B}_{-1/2}|^2 \right ] \nn \\
&&T_{010} = {1 \over 4\pi \sqrt{3}} \left [
|A_{+1/2}|^2 - |A_{-1/2}|^2 \right ] \left [ |B_{+1/2}|^2 + |B_{-1/2}|^2
\right ] \left [ |\bar{B}_{+1/2}|^2 - |\bar{B}_{-1/2}|^2 \right ] \nn \\
&&T_{110} = {1 \over 12\pi} \left [ |A_{+1/2}|^2 +
|A_{-1/2}|^2 \right ] \left [ |B_{+1/2}|^2 - |B_{-1/2}|^2 \right ] \left [
|\bar{B}_{+1/2}|^2 - |\bar{B}_{-1/2}|^2 \right ] \nn \\
&&T_{11-1} = {1 \over 6\pi}  A_{+1/2} \  A^*_{-1/2} \left [ |B_{+1/2}|^2 - 
|B_{-1/2}|^2 \right ]
\left [ |\bar{B}_{+1/2}|^2 - |\bar{B}_{-1/2}|^2 \right ]	 \quad . 
\label{21e}
\eea

\noindent Defining now amplitudes of definite CP final state for the decay $B^0_d  
\to \Lambda^+_c 
\overline{\Lambda^+_c}$  

\beq
		G_{\pm} = {1 \over \sqrt{2}} \left ( A_{+1/2} \pm  A_{-1/2} \right 
) 	\qquad 	({\rm CP} = \pm)			
\label{22e}
\eeq

\noi and amplitudes of definite parity for the decay $\Lambda^+_c  \to \Lambda 
\pi^+$

	\beq
B_{pc} = {1 \over \sqrt{2}} \left ( B_{+1/2} + B_{-1/2} \right ) \qquad  		
B_{pv} = {1 \over
\sqrt{2}} \left ( B_{+1/2} - B_{-1/2} \right ) 		 \label{23e}
\eeq

\noindent one can rewrite the moments in the form

\bea
&&T_{000} = {1 \over 4\pi}  \left [ |G_+|^2 + |G_-|^2 \right ] \left
[ |B_{pc}|^2 + |B_{pv}|^2 \right ] \left [ |\bar{B}_{pc}|^2 + \bar{B}_{pv}|^2 
\right ] \nn \\
&&T_{100} =  {1 \over 4\pi \sqrt{3}}   2{\rm Re}(G_+G^*_- ) 2{\rm 
Re}(B_{pc}B^*_{pv}
) \left [ |\bar{B}_{pc}|^2 + |\bar{B}_{pv}|^2 \right ] \nn \\
&&T_{010} = {1 \over 4\pi \sqrt{3}}   2{\rm Re}(G_+G^*_- ) \left [ |B_{pc}|^2 + 
|B_{pv}|^2 \right ]
2{\rm Re}(\bar{B}_{pc}\bar{B}^*_{pv} ) \nn \\
&&T_{110} = {1 \over 12\pi} \left [ |G_+|^2 + |G_-|^2 \right ] 2{\rm 
Re}(B_{pc}B^*_{pv} )
2{\rm Re}(\bar{B}_{pc} \bar{B}^*_{pv} ) \\
&&T_{11-1} = {1 \over 12\pi}  \left  [ |G_{+}|^2 - |G_-|^2- 2i{\rm Im}(G_+G^*_- ) 
\right ]
2{\rm Re}(B_{pc}B^*_{pv} ) 2{\rm Re}(\bar{B}_{pc}\bar{B}^*_{pv})	 \nn \quad 
.
\label{24e}
\eea

\noindent If parity were conserved in the baryon decay, the only observable would 
be $T_{000}$ or
$|G_+|^2 + |G_-|^2$ (as pointed out in ref. \cite{9r}), but since parity is in 
general violated, one
has in general Re$(B_{pc}B^*_{pv})$, Re$(\bar{B}_{pc}\bar{B}^*_{pv}) \not= 0$. In 
terms of the
parity violating parameters in the secondary decays, 

\beq
	\alpha_{\Lambda}  = {2{\rm Re}(B_{pc}B^*_{pv}) \over |B_{pc}|^2 + 
|B_{pv}|^2} \qquad  	
\bar{\alpha}_{\Lambda} = {2{\rm Re}(\bar{B}_{pc} \bar{B}^*_{pv}) \over 
|\bar{B}_{pc}|^2 +
|\bar{B}_{pv}|^2} 		 \label{25e} \eeq

\noi new moments appear due to parity violation $T_{100}$, $T_{010}$, $T_{110}$ 
and
$T_{11-1}$ res\-pec\-ti\-ve\-ly proportional to $\alpha_{\Lambda}$, 
$\bar{\alpha}_{\Lambda}$,
$\alpha_{\Lambda} \bar{\alpha}_{\Lambda}$ and $\alpha_{\Lambda} 
\bar{\alpha}_{\Lambda}$.  Neglecting
now CP violation in $\Lambda_c^+$ decay, a safe assumption in the Standard Model, 
then
$\alpha_{\Lambda} + \bar{\alpha}_{\Lambda} \cong 0$ \cite{17r}, and we see that 
the observables of
the primary decay are now

\beq
	|G_+|^2	\qquad |G_-|^2	\qquad {\rm Re}(G_+G^*_- )		\qquad 
{\rm Im}(G_+G^*_- )		
\label{26e}
\eeq

\noindent in general time dependent. CP violation in $\Lambda_c^+$ decay will be 
discussed below,
together with the present limits on CP violation in $\Lambda$ decay. Therefore we 
have more
information than in the case $J/\Psi K^*$ because of the term Re$(G_+G^*_-)$ that 
at $t = 0$ will
give $\cos [{\rm Arg} (G_+G^*_-)]$. The time dependence of the amplitudes is given 
by (assuming
$\Delta \Gamma = 0)$~:

\bea
&&G_+(t) = G_+(0) e^{-iMt} \ e^{-\Gamma t/2} \left ( \cos {\Delta Mt \over 2}  + 
i\lambda_+ \sin
{\Delta Mt \over 2} \right ) \nn \\ 
&&G_-(t) = G_-(0) e^{-iMt}\  e^{-\Gamma t/2} \left ( \cos {\Delta Mt \over 2}  - 
i\lambda_-
\sin {\Delta Mt \over 2} \right )				 \label{27e}
\eea

\noi where

\beq
			\lambda_{\pm} = {q \over p} \ {\bar{G}_{\pm}(0) \over 
G_{\pm}(0)} \quad .  							
\label{28e}
\eeq

\noi Notice that there is no general argument to claim that $\lambda_+ = 
\lambda_-$ because of the
presence of Penguins, that can differently affect the two CP amplitudes, i.e. 
their contribution
with possible FSI phases can depend on the helicity \cite{18r}. In terms of 
$\lambda_{\pm}$ the
time-dependent observables (\ref{26e}) write~:

\bea
&&|G_{\pm}(t)|^2 = e^{- \Gamma t} \left [ X_{\pm}^{(1)} + X_{\pm}^{(2)} \cos 
\Delta Mt +
X_{\pm}^{(3)} \sin \Delta Mt \right ] \nn \\ 
&&{\rm Im} \left [ G_-(t) G_+^*(t) \right ] = e^{- \Gamma t} \left [ Y^{(1)} + 
Y^{(2)} \cos \Delta Mt
+ Y^{(3)} \sin \Delta Mt \right ] \nn \\ 
&&{\rm Re} \left [ G_-(t) G_+^*(t) \right ] = e^{- \Gamma t} \left [ Z^{(1)} + 
Z^{(2)} \cos \Delta Mt
+ Z^{(3)} \sin \Delta Mt \right ] 
\label{formula1}
\eea 

\noi where

\bea
&&X_{\pm}^{(1)} = |G_{\pm}(0)|^2 {1 \over 2} \left ( 1 + |\lambda_{\pm}|^2 \right 
) \nn \\
&&X_{\pm}^{(2)} = |G_{\pm}(0)|^2 {1 \over 2} \left ( 1 - |\lambda_{\pm}|^2 \right 
) \nn \\
&&X_{\pm}^{(3)} = \mp |G_{\pm}(0)|^2 {\rm Im} \lambda_{\pm} 
\label{formula2}
\eea

\noi and

\bea
&&Y^{(1)} = {\rm Re}\left [ G_-(0) G_+^*(0) \right ]{1 \over 2} \left ( {\rm Re} 
\lambda_- {\rm Im}
\lambda_+ - {\rm Re} \lambda_+ {\rm Im} \lambda_- \right ) \nn \\ 
&&+ {\rm Im}\left [ G_-(0) G_+^*(0) \right ] {1 \over 2} \left ( 1 - {\rm Im} 
\lambda_- {\rm Im}
\lambda_+ - {\rm Re} \lambda_- {\rm Re}\lambda_+ \right ) \nn \\
&& \nn \\
&&Y^{(2)} = - {\rm Re}\left [ G_-(0) G_+^*(0) \right ] {1 \over 2} \left ( {\rm 
Re} \lambda_- {\rm
Im} \lambda_+ - {\rm Re} \lambda_+ {\rm Im} \lambda_- \right ) \nn \\ 
&&+ {\rm Im}\left [ G_-(0) G_+^*(0) \right ] {1 \over 2} \left ( 1 + {\rm Im} 
\lambda_- {\rm Im}
\lambda_+ + {\rm Re} \lambda_- {\rm Re}\lambda_+ \right ) \nn \\
&& \nn \\
&&Y^{(3)} = - {\rm Re}\left [ G_-(0) G_+^*(0) \right ] {1 \over 2} \left ( {\rm 
Re} \lambda_- + {\rm
Re} \lambda_+ \right ) \nn \\
&&+ {\rm Im} \left [  G_-(0)G_+^*(0) \right ] {1 \over 2} \left ( {\rm Im}
\lambda_- - {\rm Im} \lambda_+ \right ) \nn \\
&& \nn \\
&&Z^{(1)} = {\rm Re}\left [  G_-(0) G_+^*(0) \right ] {1 \over 2} \left ( 1 - {\rm 
Re} \lambda_+ {\rm
Re} \lambda_- - {\rm Im} \lambda_+ {\rm Im} \lambda_- \right ) \nn \\ 
&&- {\rm Im}\left [  G_-(0) G_+^*(0) \right ] {1 \over 2} \left ( {\rm Im} 
\lambda_+ {\rm Re}
\lambda_- - {\rm Re} \lambda_+ {\rm Im}\lambda_- \right ) \nn \\
&& \nn \\
&&Z^{(2)} = {\rm Re}\left [ G_-(0) G_+^*(0) \right ] {1 \over 2} \left ( 1 + {\rm 
Re} \lambda_+ {\rm
Re} \lambda_- + {\rm Im} \lambda_+ {\rm Im} \lambda_- \right ) \nn \\ 
&&+ {\rm Im}\left [ G_-(0) G_+^*(0) \right ] {1 \over 2} \left ( {\rm Im} 
\lambda_+ {\rm Re}
\lambda_- - {\rm Re} \lambda_+ {\rm Im}\lambda_- \right ) \nn \\
&& \nn \\
&&Z^{(3)} = - {\rm Re}\left [ G_-(0) G_+^*(0) \right ] {1 \over 2} \left ( {\rm 
Im} \lambda_+ - {\rm
Im} \lambda_- \right ) \nn \\
&&+ {\rm Im} \left [ G_-(0) G_+^*(0) \right ] {1 \over 2} \left ( {\rm Re}
\lambda_+  + {\rm Re}\lambda_- \right ) \ . 
\label{formula3}
\eea

\noindent Hopefully separating the different time dependences $e^{-\Gamma t}$, 
$e^{-\Gamma t} \cos
\Delta Mt$ and $e^{-\Gamma t} \sin \Delta Mt$, all the quantities $X_{\pm}^{(i)}$, 
$Y^{(i)}$ and
$Z^{(i)}$ $(i = 1, 2, 3)$ are in principle observable. We have four unknowns 
related to CP violation,
namely Im$\lambda_+$, Im$\lambda_-$ and Re$\lambda_+$, Re$\lambda_-$. As usual, 
the imaginary parts
Im$\lambda_+$, Im$\lambda_-$ can be measured by the different terms contributing 
to $|G_{\pm}(t)|^2$
(\ref{formula2}) since $|\lambda_{\pm}|^2$ can also be known. Moreover, we have 
two strong
interaction unknowns, namely Re$[G_-(0)G_+^*(0)]$ and Im$[G_-(0)G_+^*(0)]$. As we 
now show, the other
four unknowns Re$\lambda_+$, Re$\lambda_-$ and Re$[G_-(0)G_+^*(0)]$, 
Im$[G_-(0)G_+^*(0)]$ can be
determined without discrete ambiguities by the six observables (\ref{formula3}). 
First, one finds~:

\begin{eqnarray}
&&{\rm Im} \left [ G_-(0) G_+^*(0) \right ] =   Y^{(1)} + Y^{(2)}  \nn \\
&&{\rm Re} \left [  G_-(0) G_+^*(0) \right ] =   Z^{(1)} + Z^{(2)}  \quad .
\label{formula4}
\end{eqnarray}

\noindent Calling now the combinations of observables

\begin{eqnarray}
&&r = {{\rm Re} \left [ G_-(0)G_+^*(0) \right ] \over {\rm Im} \left [ G_-(0) 
G_+^*(0) \right ]} \nn
\\ &&A = {r \over 1 + r^2}  \left [ {Y^{(1)} -
Y^{(2)} \over Y^{(1)} + Y^{(2)}} - {Z^{(1)} - Z^{(2)} \over Z^{(1)} + Z^{(2)}} 
\right ] \nn \\
&&B = {2r \over 1 + r^2} \left [ {Z^{(3)} \over Z^{(1)} + Z^{(2)}} - {Y^{(3)} 
\over Y^{(1)} +
Y^{(2)}} \right ]
\label{formula5}
\end{eqnarray}

\noindent we find

\begin{equation}
\label{formula6}
{\rm Re} \lambda_+ = - {A - B{\rm Im} \lambda_+ \over {\rm Im} \lambda_+ + {\rm
Im} \lambda_-} \qquad {\rm Re} \lambda_- = {A + B{\rm Im} \lambda_- \over {\rm
Im} \lambda_+ + {\rm Im} \lambda_-} \quad .
\end{equation}

The conclusion is that the six unknowns Im$\lambda_+$, Im$\lambda_-$, 
Re$\lambda_+$, Re$\lambda_-$,
Re$[G_-(0)G_+^*(0)]$ and Im$[G_-(0)G_+^*(0)]$ can be determined in principle 
without discrete
ambiguities, and the system is \underbar{overconstrained} since we have 12 
observables. \\

\section{Particular cases}
\hspace*{\parindent} Let us first examine the case $B_d^0(t) \to \Lambda_c^+
\overline{\Lambda_c^+}$, related to the angle $\beta$ (Fig. 2). The decay 
amplitude will write, in
all generality

\beq
G_{\pm} = G_{\pm}^{(u)} V_{ud}^* V_{ub} + G_{\pm}^{(c)} V_{cd}^* V_{cb} + 
G_{\pm}^{(t)} V_{td}^*
V_{tb} 
\label{28bis}
\eeq

\noindent where $G_{\pm}^{(c)}$ and $G_{\pm}^{(t)}$ are respectively the 
current-current (tree) and
short distance Penguin amplitudes, and $G_{\pm}^{(u)}$ is a long distance 
$u$-Penguin, responsible
for rescattering effects \cite{18bisr}. Using unitarity $V^*_{ud} V_{ub} + 
V^*_{cd} V_{cb} + V^*_{td}
V_{tb} = 0$, \underbar{in all} \underbar{generality} $\lambda_{\pm}$  will be of 
the form

\beq
			\lambda_{\pm} = \pm  {e^{-i\beta} - |R_{\pm}| 
e^{i\delta_{\pm}} \over e^{i\beta} -
|R_{\pm}| e^{i\delta_{\pm}}} 					 \label{29e}
\eeq

\noi where 

\beq
	R_{\pm} = z r_{\pm} \qquad 		z = {V^*_{td}V_{tb} \over 
V^*_{cd}V_{cb}} 	\qquad 	r_{\pm} =
{P_{\pm} \over T_{\pm}}  = |r_{\pm}|e^{i\delta_{\pm}}		 \label{30e}	
\eeq

\noi $z$ being a ratio of CKM matrix elements, and $r_{\pm}$ the ratio of strong
amplitudes~: 

\beq
\label{30bis}
P_{\pm} = G_{\pm}^{(t)} - G_{\pm}^{(u)} \quad , \quad T_{\pm} = G_{\pm}^{(c)} - 
G_{\pm}^{(u)} \quad
. \eeq

\noi The notation $P_{\pm}$, $T_{\pm}$ means that these are respectively 
Penguin and dominantly tree amplitudes. One can measure in principle 
Re$\lambda_{\pm}$ and
Im$\lambda_{\pm}$ and we have 5
unknowns, namely $\beta$, $|R_{\pm}|$ and $\delta_{\pm}$. Therefore, to get 
information on $\beta$
\underbar{we need theoretical input on a single parameter}. This is the general 
situation in these
decays if Penguins are sizeable. At this stage, one may use the formalism of 
\cite{5bisr}~: one
can get $\beta$ as a function of this unknown parameter and of the observables. 
Then, measuring
Re$\left ( {q \over p} {\bar{G} \over G} \right )$ means that a discrete ambiguity 
$(\beta_{eff}
\to {\pi \over 2} - \beta_{eff})$ is solved in this procedure, and allows to get 
more information
than in, e.g., $B_d^0 \to D^+D^-$, \underbar{even in the presence of sizeable 
Penguin}\break
\noindent \underbar{uncertainties}. 

Notice that $|z| \sim O(1)$ in powers of the Wolfenstein parameter
$\lambda$, although presumably Penguins are not large because of their small short 
distance
coefficient, or loop suppression for long distance Penguins. \underbar{If we 
assume that the}\break
\noindent \underbar{Penguin contribution is small}, i.e. $R_{\pm}$ is small, the 
time dependence of
the observables is given in terms of $\beta$ as follows~: 
\bea
	\label{31e}
&&|G_+(t)|^2 = |G_+(0)|^2 \ e^{-\Gamma t} \left ( 1 + \sin2 \beta \sin \Delta Mt 
\right ) \nn \\
&& \nn \\
&&|G_-(t)|^2 = |G_-(0)|^2 \ e^{-\Gamma t} \left ( 1 - \sin 2\beta \sin \Delta Mt 
\right)		\nn \\
&& \nn \\		
&&{\rm Im}\left [ G_-(t)G^*_+(t) \right ] = e^{-\Gamma t} \Big \{ {\rm Im}\left [ 
G_-(0)G^*_+(0)
\right ]  \cos \Delta Mt - \nn \\ &&\qquad {\rm Re}\left [ G_-(0)G^*_+(0) \right ] 
\cos 2 \beta \sin
\Delta Mt \Big \} \nn \\ 
&& \nn \\
&&{\rm Re}\left [ G_-(t)G^*_+(t) \right ]  = e^{-\Gamma t} \Big \{ {\rm Re}\left [ 
G_-(0)G^*_+(0)
\right ]  \cos \Delta Mt + \nn \\ &&\qquad {\rm Im}\left [ G_-(0)G^*_+(0) \right ] 
 \cos 2\beta \sin
\Delta Mt \Big \}  \quad . \eea

\noi One can in principle separate both Im$[G_-(t)G^*_+(t)]$  and 
Re$[G_-(t)G^*_+(t)]$  by the
angular analysis. The term in $e^{-\Gamma t} \sin \Delta Mt$ from 
Im$[G_-(t)G^*_+(t)]$ measures
sign$(\cos \delta \cos 2\beta )$ where $\delta = {\rm Arg}[G_-(0)G^*_+(0) ]$, but 
the sign$(\cos
\delta )$ can be known from the coefficient of $e^{-\Gamma t} \cos \Delta Mt$ in 
the new observable
Re$[G_-(t)$ $G^*_+(t)]$. Conversely, one could measure sign$(\sin \delta \cos 
2\beta)$ from the term
in $e^{-\Gamma t} \sin \Delta Mt$ from Re$[G_-(t)G^*_+(t)]$, and sign$(\sin 
\delta)$ can be known
from the coefficient of $e^{-\Gamma t} \cos \Delta Mt$ in the observable 
Im$[G_-(t)G^*_+(t)]$.
Therefore, $\cos 2\beta$ is overdeterminated in this case in which parity is 
violated in the
secondary decay.
	
However, in the case of $\beta$, this procedure could be quite difficult to put in 
practice, since,
as we have seen above, the branching ratio of $B^0_d  \to \Lambda^+_c 
\overline{\Lambda^+_c}$ being
Cabibbo-suppressed can be expected to be of the order $10^{-4}$ and the exclusive
$\Lambda^+_c$ two-body decays like $\Lambda^+_c  \to \Lambda \pi^+$, $p\bar{K}^0$, 
$\Sigma^0 \pi^+$,
... have already been measured with BR of the order of $10^{-2}$. Notice also that 
the parity
violation $\alpha$ parameter of some of these modes has already been measured~: 
$\alpha (\Lambda
\pi^+) = - 0.98 \pm 0.19$, $\alpha (\Sigma^0\pi^+) = -0.45 \pm 0.32$. Since the 
combined branching
ratios in the case of $\beta$ are small, one should look for inclusive arguments 
along the same
lines, in order to increase statistics. This might be possible to carry out, since 
the solution of
the discrete ambiguity $\beta \to {\pi \over 2}  - \beta$ only needs the 
determination of sign($\cos
2\beta )$.  
	
Let us now turn to one major purpose of this paper, the determination of $\sin 
2\alpha$
and $\cos 2\alpha$ (up to Penguins) in the sequential decay $B^0_d(t) \to \Lambda 
\bar{\Lambda}$,
$\Lambda  \to p\pi^-$. The time dependence of the amplitudes is given by the same 
expressions
(\ref{27e}) with $\lambda_{\pm}$ now given by

\beq
			\lambda_{\pm} = \pm  {e^{i\alpha} - |R_{\pm}| 
e^{i\delta_{\pm}} \over e^{-i\alpha} -
|R_{\pm}|e^{i\delta_{\pm}}} 					 \label{32e}
\eeq

\noi where 

\beq
	R_{\pm} = z r_{\pm}		\qquad z = {V^*_{td} V_{tb} \over V^*_{ud} 
V_{ub}} 	\qquad 	r_{\pm} =
{P_{\pm} \over T_{\pm}}  = |r_{\pm}|e^{i\delta_{\pm}}		 \label{33e}
\eeq

\noi z being a ratio of CKM matrix elements, and $r_{\pm}$ is the Penguin to tree 
ratio of strong
amplitudes, different than in the $\beta$ case (\ref{30e}), but defined along the 
same lines. Since
$z \cong {1-\rho +i \eta \over \rho -i \eta}$  , there is no CKM suppression of 
Penguins. From the
values of the short distance QCD coefficients, one naively expects $|R_{\pm}| 
\cong 0.05 \times
\left | {1 - \rho + i \eta \over \rho - i \eta} \right | \sim 0.15$, not 
inconsistent with CLEO data
\cite{2r}. \underbar{Neglecting Penguins}, the observables will read~:

\bea
&&|G_+(t)|^2 = |G_+(0)|^2 \ e^{-\Gamma t} \left (1 - \sin 2\alpha \sin \Delta Mt 
\right ) \nn \\
&&|G_-(t)|^2 = |G_-(0)|^2 \ e^{-\Gamma t} \left ( 1 + \sin 2\alpha \sin \Delta Mt 
\right )		\nn	\\		
&&{\rm Im}\left [ G_-(t)G^*_+(t) \right ] = e^{-\Gamma t} \Big \{ {\rm Im}\left [ 
G_-(0)G^*_+(0)
\right ]  \cos \Delta Mt - \nn \\ &&\qquad {\rm Re}\left [ G_-(0)G^*_+(0) \right ] 
\cos 2\alpha \sin
\Delta Mt \Big \} \nn \\ &&{\rm Re}\left [  G_-(t)G^*_+(t) \right ] = e^{-\Gamma 
t} \Big \{ {\rm
Re}\left [ G_-(0)G^*_+(0) \right ] \cos \Delta Mt + \nn \\ &&\qquad {\rm Im} \left 
[
G_-(0)G^*_+(0) \right ]  \cos 2\alpha \sin \Delta Mt \Big \} \quad . \label{34e} 
\eea

\noindent As pointed out above, in view of the intrinsic difficulties in the
determination of $\alpha$ with any decay mode, we are here in a relatively 
favorable situation. One
expects a $BR(B^0_d  \to \Lambda \bar{\Lambda}) \sim 10^{-5}-10^{-6}$, one has an 
excellent detection
efficiency for $\Lambda  \to p\pi^-$ with a large $BR(\Lambda \to p\pi^-) = 64 \ 
\%$, and a sizeable
parity violation parameter $\alpha_{\Lambda} (p\pi^-) = 0.64$. Notice however that 
in order to
extract the observables (\ref{34e}) one needs to neglect CP violation in $\Lambda$ 
decay \cite{17r},
which, in the Standard Model, would come from CKM suppressed Penguin diagrams. 
Anyhow, to extract
(\ref{34e}) one needs to assume  that the CP violation parameter $\alpha + 
\bar{\alpha}$ is very
small. Taking into account the expected size of CP violation in $B$ decays, 
$\alpha_{\Lambda} +
\bar{\alpha}_{\Lambda} \cong 0$ is a safe assumption, since present data give for 
the CP violation
parameter in $\Lambda$ decay \cite{7r}~:

\beq
\label{34bis}
{\alpha_{\Lambda} + \bar{\alpha}_{\Lambda} \over \alpha_{\Lambda} - 
\bar{\alpha}_{\Lambda}} = - 0.03
\pm 0.06 \ \ \ . \eeq

\noi The Standard Model predicts a tiny value of $O(10^{-5})$ \cite{17r} due to 
Penguins. Penguin
diagrams being much smaller in charm decay, the Standard Model prediction of CP 
violation for
$\Lambda_c^+$ decay is even smaller. Another difficulty of quite a different 
nature is
that since the $\Lambda$ decays far away from the primary vertex, it may be 
experimentally
difficult to measure the time dependence of the observables \cite{19r}. 
	
Let us now turn to the possible determination of $\sin 2\gamma$ and $\cos 2\gamma$ 
through the decays
$B^0_s(t) \to \Lambda \bar{\Lambda}$, $\Xi^0\overline{\Xi^0}$. The time dependence 
of
the amplitudes is given by the same expressions (\ref{27e}) (making the rough 
approximation of
neglecting $(\Delta \Gamma )_{B_s}$) with now $\lambda_{\pm}$ given now by

\beq
			\lambda_{\pm} = \pm  {e^{-i\gamma} 
-|R_{\pm}|e^{i\delta_{\pm}} \over e^{i\gamma}
-|R_{\pm}|e^{i\delta_{\pm}}} 					 \label{35e}
\eeq

\noindent where 

\beq
	R_{\pm} = z r_{\pm}		\qquad z = {V^*_{ts}V_{tb} \over 
V^*_{us}V_{ub}} \qquad 		r_{\pm} =
{P_{\pm} \over T_{\pm}}  = |r_{\pm}|e^{i\delta_{\pm}}		 \label{36e}
\eeq

\noindent $z$ being a ratio of CKM matrix elements, and $r_{\pm}$ the ratio of 
dominantly Penguin to
tree strong amplitudes, different than in the $\beta$ and $\alpha$ cases 
(\ref{30e}) and
(\ref{36e}). Since now we have $z \cong {1 \over \lambda^2(\rho -i \eta)}$, there 
is CKM enhancement
of the Penguin to tree ratio, that could compensate the ratio $|r_{\pm}| \sim 
0.05$,
giving $|R_{\pm}|\sim {|r_{\pm}| \over \lambda^2} {1 \over |\rho - i \eta|} \sim 
3$. To conclude,
branching ratios are very small $\sim 10^{-7} - 10^{-8}$, and Penguins could be 
relatively quite
important in this case due to the tree suppression. These modes do not seem 
suitable to get
information on $\gamma$.   

Let us now comment about the pure Penguin modes $B^0_d(t)$ or $B^0_s(t) \to
\Sigma^-\overline{\Sigma^-}$, $\Xi^-\overline{\Xi^-}$, 
$\Omega^-\overline{\Omega^-}$. If short
distance Penguin dominates, the weak phase of ${q \over p}$ cancels the
weak phase of the ratio of decay amplitudes ${\bar{G}_f \over G_f}$ for 
\underbar{both} $B_d$ or
$B_s$, so that the product $\lambda_f = {q \over p} {\bar{G}_f \over G_f}$ is 
real. However, long
distance Penguins induced by $u$ or $c$ quark loops could contribute also. These 
terms
describe rescattering effects of the form $B^0_q(t) \to (\bar{u}q)(u\bar{q})\to
\Sigma^-\overline{\Sigma^-}$, $\Xi^-\overline{\Xi^-}$, 
$\Omega^-\overline{\Omega^-}$ ($q = d$ or
$s$) . 

Let us first consider $B^0_d(t) \to \Sigma^-\bar{\Sigma}^-$, $\Xi^-\bar{\Xi}^-$,
$\Omega^-\bar{\Omega}^-$. In this case, in the Standard Model, the time dependence 
of the amplitudes
is given by the same expressions (\ref{27e}) with now $\lambda_{\pm}$  given now 
by

\beq
			\lambda_{\pm} = \pm  {1- |R_{\pm}| e^{i\alpha} \ 
e^{i\delta_{\pm}} \over 1-
|R_{\pm}| e^{-i\alpha} \ e^{i\delta_{\pm}}} 					 
\label{37e}
\eeq

\noindent where now

\beq
	R_{\pm} = z r_{\pm}	 	\qquad z = {V^*_{ud} V_{ub} \over V^*_{td} 
V_{tb}} \qquad 		r_{\pm} =
{P^{LD}_{\pm} \over P_{\pm}}  = |r_{\pm}|e^{i\delta_{\pm}}	 \label{38e}
\eeq

\noindent where $r_{\pm}$ is now the ratio of $P_{\pm}^{LD}$ (the difference 
between long distance
$u$ and $c$ Penguins) to $P_{\pm}$ (difference between $t$ and $c$ Penguins) 
strong amplitudes.
Since $z \cong {\rho -i\eta \over 1-\rho +i\eta}$, $|z| \sim 0.36$, there is no 
CKM suppression of
$LD$ Penguins. Neglecting however these rescattering effects, that might be small,
 $\lambda_{\pm}$ will be real and these modes can be useful to get information on
possible sources of CP violation beyond the Standard Model. However, branching 
ratios are quite
unfavorable in this case, as pointed out above. The situation is  better for 
$B^0_s(t) \to
\Sigma^-\overline{\Sigma^-}$, $\Xi^-\overline{\Xi^-}$, 
$\Omega^-\overline{\Omega^-}$, for which
$\lambda_{\pm}$ is given by

\beq
			\lambda_{\pm} = \pm  {1- |R_{\pm}| e^{-i\gamma} \ 
e^{i\delta_{\pm}} \over 1-
|R_{\pm}|e^{i\gamma} \ e^{i\delta_{\pm}}} 					 
\label{39e}
\eeq

\noindent where

\beq
	R_{\pm} = z r_{\pm}	 	\qquad z = {V^*_{us}V_{ub} \over 
V^*_{ts}V_{tb}} 	\qquad 	r_{\pm} =
{P^{LD}_{\pm} \over P_{\pm}}  = |r_{\pm}|e^{i\delta_{\pm}} \quad .	 
\label{40e}
\eeq

\noindent Since now $z \cong - \lambda^2(\rho - i\eta )$ is CKM suppressed, we can 
safely
\underbar{neglect long} \underbar{distance Penguins} and $\lambda_{\pm}$ is 
predicted to be real in
the Standard Model. Analogously to $B_d^0 \to \varphi K_S$, these modes can then 
be useful to get
information on possible sources of CP violation beyond the Standard Model, i.e., 
to obtain
Im$\lambda_{\pm}$  and Re$\lambda_{\pm}$ whatever the origin of CP violation could 
be. \\

\section{CP violation in $\Lambda$ decay from B$_{\bf d}^{\bf 0} \to \Lambda 
\bar{\Lambda}$}
\hspace*{\parindent} 
As a quite different application of the formalism developed here, let us consider 
CP violation in
$\Lambda$ decay. A simple inspection of (26) shows that the CP violation parameter 
in
$\Lambda$ decay is given by

\beq
\label{42bis}
{\alpha_{\Lambda} + \bar{\alpha}_{\Lambda} \over \alpha_{\Lambda} - 
\bar{\alpha}_{\Lambda}} =
{T_{100} + T_{010} \over T_{100} - T_{010}} \quad . \eeq

\noi As we now discuss, this quantity is, in general, \underbar{independent} of CP 
violation in $B_d$
decay, and the formula holds also for \underbar{time-integrated} moments. This can 
be interesting
since the $B_d^0$ is a coherent source of $\Lambda\bar{\Lambda}$ and provides 
automatically a
callibration of the measure of $\alpha_{\Lambda}$ and $\bar{\alpha}_{\Lambda}$. 

Formulas (26) show that $T_{100}$ and $T_{010}$ are proportional to 
Re$[G_+(t)G_-^*(t)]$. In
these expressions, there are three kinds of time dependence~:

\beq
\label{for1}
e^{\Gamma t} {1 \over 2} {\rm Re}\left [ G_+(0) G_-^*(0) - \bar{G}_+(0) 
\bar{G}_-^*(0) \right ]
\eeq

\beq
\label{for2}
e^{\Gamma t} \cos \Delta Mt {1 \over 2} {\rm Re}\left [ G_+(0) G_-^*(0) + 
\bar{G}_+(0)
\bar{G}_-^*(0) \right ] \eeq

\beq
\label{for3}
e^{\Gamma t} \sin \Delta Mt {1 \over 2} {\rm Im}\left [ {q \over p} \bar{G}_+(0) 
G_-^*(0) + {p
\over q} G_+(0) \bar{G}_-^*(0) \right ] \quad .
\eeq 

\noi We have assumed in these expressions that ${q \over p}$ is a pure phase, as 
given in the
Standard Model, and coherent with $\Delta \Gamma = 0$. Here the $\bar{G}_{\pm}(0)$ 
amplitudes are
defined from the CP transformation of $G_{\pm}(0)$ by the relation 
$\bar{G}_{\pm}(0) = \pm CP
[G_{\pm}(0)]$. The first term (\ref{for1}) corresponds to direct CP violation, 
because it vanishes if
Penguins are absent. This term survives for untagged events and obviously it 
survives also if one
integrates over time. The second term (\ref{for2}) conserves CP. It survives 
integrating over time
but does not contribute to untagged events. Note the interesting point that, 
\underbar{just opposite
to the usual case}, e.g. $B_d^0 \to \pi^+\pi^-$, the terms in $e^{-\Gamma t}$ and 
$e^{- \Gamma t}
\cos \Delta Mt$ in these angular correlations correspond to direct CP violation 
and to CP
conservation. This is due to the fact that we have here the real part of the 
interference between two
\underbar{opposite CP} amplitudes. Also unusual is the role of the third term 
(\ref{for3}) that in
the present case conserves CP, and is sensitive to $\cos 2 \Phi$, $\Phi$ being a 
CP angle. This
term  survives in general the time integration, except in $e^+e^-$ at the 
$\Upsilon (4S)$, although
tagging is necessary. \par

To conclude, formula (\ref{42bis}) holds in general, \underbar{even if one 
integrates over time}.
There is one exception, namely the case of untagged events in the limit of 
vanishing Penguin.
Indeed, in this last case both $T_{100}$ and $T_{010}$ vanish. Of course, the 
possibility of
integrating over time is very welcome although tagging is needed in practice. 
However, using
$B_d^0 \to \Lambda \bar{\Lambda}$ would need large statistics, maybe available in 
hadronic
machines. In view of the present accuracy (45), it would be suitable to reach at 
least a 1~\%
upper limit on the $\Lambda$ CP asymmetry. To this aim, one would need of the 
order of $10^{10}$
$B_d^0$ mesons, since this number is inversely proportional to the $BR(B_d^0 \to 
\Lambda
\bar{\Lambda}) \sim 10^{-6}$ and to the square of the CP asymmetry.  \\

\section{Conclusion}
\hspace*{\parindent} 
A matter of principle is the main point of this paper, namely that not only
$\sin 2 \beta$ ($\sin 2 \alpha$) but also $\cos 2 \beta$ ($\cos 2 \alpha$) (up to 
Penguin
pollution) could be reached in decays of the type $B^0 \to$ hyperon + antihyperon, 
because of
parity violation in the hyperon decay, provided one neglects CP violation in the 
latter, a safe
assumption within the Standard Model. On the other hand, we have also shown that 
this decay
can be useful to look for CP violation in hyperon decay. However, one should keep 
in mind the
smallness of the needed combined branching ratios and, also, in the 
in\-te\-res\-ting case
$\Lambda\bar{\Lambda}$, that the time-dependence of the decay could be hard to 
measure because
the $\Lambda$ decays far away from the primary vertex. \\

\section*{Acknowledgement}
\hspace*{\parindent}
We are indebted to Y. Grosman and H. Quinn for pointing out to us that in the 
$B^0_d(t)
\to J/\Psi K^{*0} \to \ell^+\ell^-K_S\pi^0$ angular correlations one does not 
measure $\cos 2\beta$
but $\cos \delta \cos 2\beta$, $\delta$ being a strong phase. We also acknowledge 
useful remarks
from B. D'Almagne, A. Gaidot and G. Vasseur. \par

The authors acknowledge partial support from the EEC-TMR Program, Contract N. 
CT98-0169. \\

\section*{Note added}
\hspace*{\parindent}
When this work was finished, we noticed two recent papers by A. S. Dighe 
\underbar{et al.}
\cite{21ters} that discuss the determination of sign($\cos 2 \beta$) using the 
decays $B_{u,d} \to
J/\psi K^*$ and $B_s \to J/\psi \varphi$. The sign($\cos 2 \beta )$ could be 
determined using
SU(3) and interference effects involving the sizeable $\Delta \Gamma$ of the 
$B_s$-$\bar{B}_s$
system, which is neglected in our paper.

\newpage
\section*{Figure Captions}

\noi {\bf Fig. 1 :} \par
Decay $B_d^0 \to \Lambda \bar{\Lambda}$. The four-fermion interaction represents 
local operators,
current-current (related to the CP angle $\alpha$), or Penguins. \\

\noi {\bf Fig. 2 :} \par
Decay $B_d^0 \to \Lambda_c^+ \overline{\Lambda_c^+}$. The four-fermion interaction 
represents local operators,
current-current (related to the CP angle $\beta$), or Penguins. \\

\noi {\bf Fig. 3 :} \par
Decay $B_s^0 \to \Lambda\bar{\Lambda}$. The four-fermion interaction represents 
local operators, current-current
(related to the CP angle $\gamma$), or Penguins. \\

\noi {\bf Fig. 4 :} \par
Decay $B_s^0 \to \Sigma^-\overline{\Sigma^-}$. The four-fermion interaction 
represents Penguin local operators.


\begin{thebibliography}{99}
\bibitem{12r} We will adopt this somewhat unusual notation to make easier the 
comparison with the
$\Lambda^+_c \overline{\Lambda^+_c}$  mode.
\end{thebibliography}
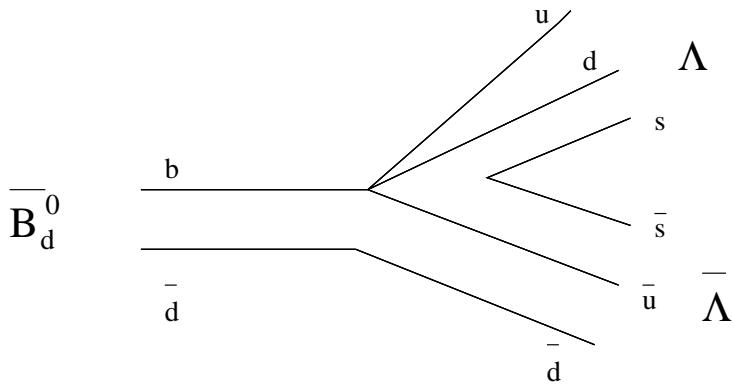
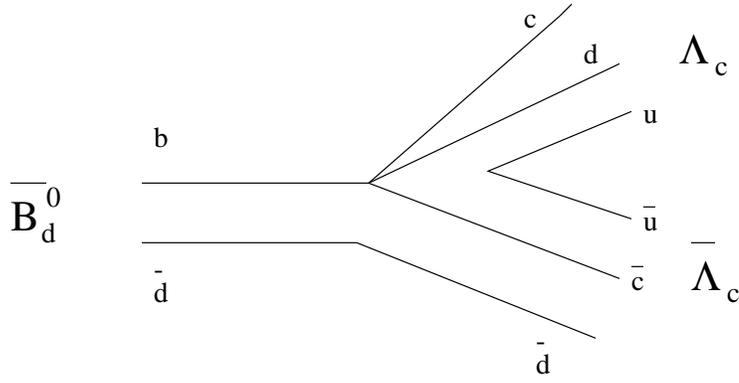
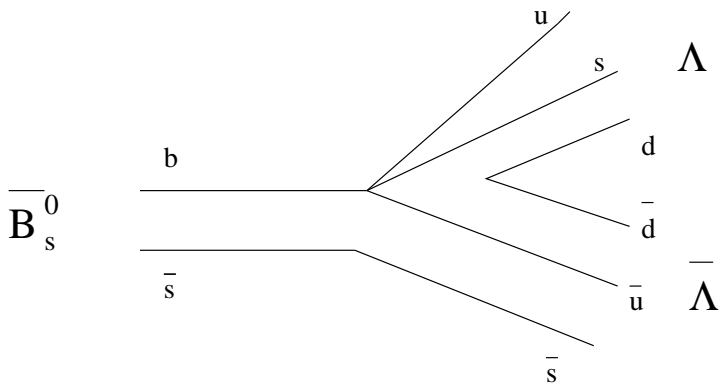
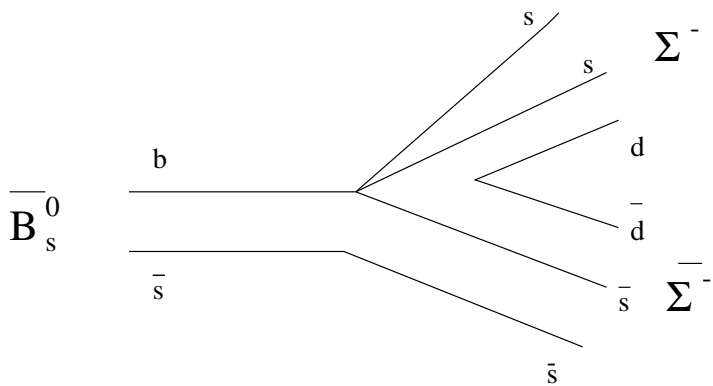
\end{document}